\documentclass[%
 reprint,
 superscriptaddress,
 amsmath,amssymb,
 aps,
 prl,
floatfix,
]{revtex4-2}

\newcommand{\partder}[2]{\frac{\partial #1}{\partial #2}}
\usepackage{graphicx}
\usepackage{dcolumn}
\usepackage{bm}
\usepackage{xcolor}
\newcommand{\hd}[1]{\textit{#1}---}

\def\arxivversion{1}

\begin{document}

\title{Nonlinear kinetic closures for linear instabilities} 

\author{Elizabeth J. Paul}
\email{ejp2170@columbia.edu}
\affiliation{Columbia University}

\author{Archis Joglekar}
\affiliation{Ergodic LLC}

\author{Alexey Knyazev}
\affiliation{Columbia University}

\date{\today}

\begin{abstract}
Landau fluid closures are linear in the evolved moments. We examine the consequences of relaxing linearity. As an example, we introduce the nonlinear exact kinetic response closure, exact on an eigenmode but not on superpositions.
A counting argument shows that a closure on $N$ instantaneous moments and the field can represent at most $(N+1)/2$ superposed eigenmodes. Guided by this, we train an equivariant neural-network and a learned Pad\'{e} closure on analytic data.
For bump-on-tail and slab ITG instabilities they reduce growth-rate errors by up to three orders of magnitude over Landau closures.
\end{abstract}

\maketitle



Kinetic phenomena such as Landau damping, kinetic instabilities, and phase mixing are critical to modeling plasma systems, yet fluid models remain essential given their relative simplicity.
Since the evolution equation for each moment involves the next, closing the hierarchy requires a prescription for the highest moment in terms of the lower ones. Kinetic effects can be retained with Landau fluid closures, which close the hierarchy through asymptotic matching of the kinetic response function in large and small argument limits. Their accuracy has recently been improved by retaining more poles in the rational approximation \cite{hunana2018new,hunana2019introductory} and by introducing wavenumber-dependent coefficients \cite{sun2026wave}. 

Landau fluid closures are widely applied: to energetic-particle driven MHD modes in magnetic confinement devices, as in the FAR3D model \cite{varela2024stability,hedrick1992alpha,spong2013simulation}, to edge \cite{chen2019extension,pitzal2023landau} and core \cite{staebler2007theory,kinsey2008first} turbulence simulations of tokamak plasmas, to magnetic reconnection \cite{allmann2018temperature}, and to global simulations of the magnetosphere \cite{ng2020improved}. Further improvement in closure models could therefore have far-reaching impact throughout plasma modeling. 

The accuracy of Landau closure models generally degrades for intermediate arguments, and is ultimately limited by the number of evolved fluid moments. A critical assumption of the Landau closure is linearity of the highest moment with respect to the lower moments. While this enables a rational approximation of the response function, it also limits the closure's fidelity. We instead relax this assumption and introduce an analytic closure, the exact kinetic response (EKR) closure, which reproduces the exact linear kinetic response of an eigenmode at its complex frequency while closing at the temperature moment. 

Recently, data-driven closures have been developed for plasma fluid models, including neural-network (NN) closures trained on kinetic simulation data \cite{joglekar2023machine,cheng2023data,Huang2025MLHeatFluxClosure,barbour2025machine} and sparse-regression methods that recover closure terms from first-principles simulations \cite{alves2022data,donaghy2023search,ingelsten2025data} (see \cite{burles2025machine} for a comprehensive review). However, these models face several obstacles to routine use in fluid simulations. First, generating training data from kinetic simulations is expensive, so closures are often trained on a few simulations at fixed parameters \cite{Huang2025MLHeatFluxClosure, wei2023data}, limiting generalization. Second, closures are frequently validated only offline against kinetic data, without deployment in a fluid simulation \cite{wei2023data,miloshevich2026electron}. 
Third, a black-box NN closure respects neither the symmetries of linear physics nor causality, and may introduce spurious instabilities.
Motivated by the EKR closure, we address these obstacles with a data-generation technique based on the linearized moment hierarchy rather than kinetic simulations. In this way, the training data is exact, inexpensive, and densely samples the complex frequency plane, including strongly-damped modes that are difficult to extract from kinetic simulations. 
Furthermore, the moment hierarchy yields a counting argument that constrains any closure acting on instantaneous moments and the field: representing $M$ superposed eigenmodes requires evolving $N = 2M-1$ moments. We learn two complementary closures from the same data: an equivariant NN closure, whose architecture enforces the symmetries of linear physics and captures superpositions of eigenmodes, and a learned Pad\'{e} closure within the Landau fluid family, whose coefficients are fit over the mode region of interest rather than by asymptotic matching.
We demonstrate these closures on a 1D-1V Vlasov-Poisson bump-on-tail instability \cite{Berk1995}, a prototypical model for the wave-particle interactions addressed by FAR3D, and on the slab ITG instability, for which the Hammett-Perkins closure was developed \cite{hammett1990fluid}.




\hd{Exact kinetic response closures}We begin with the linear 1D-1V Vlasov-Poisson equation for species $s$:
\begin{align}
    \partder{f_{1s}}{t} + v \partder{f_{1s}}{x} + \frac{q_s E}{m_s} \partder{f_{0s}}{v} = 0,
    \label{eq:Vlasov}
\end{align}
for arbitrary equilibrium distribution function $f_{0s}(v)$, characterized by density $n_{s} = \int dv \, f_{0s}(v)$, flow $u_{s} = n_s^{-1}\int dv \, vf_{0s}(v)$, and thermal spread $v_{ts}^2 = 2n_s^{-1}\int dv \, (v - u_{s})^2 f_{0s}(v)$. We define the normalized moments of $f_{1s}$ and $f_{0s}$ as $U_{n,s} = v_{ts} n_{s}^{-1} \int dw \, w^n f_{1s}$ and $M_{n,s} = v_{ts} n_{s}^{-1} \int dw \, w^n f_{0s}$, respectively, where $w = (v-u_{s})/v_{ts}$. After Fourier transforming in space and time, $f_{1s}, \Phi \sim e^{i(kx-\omega t)}$, the corresponding moment hierarchy reads,
\begin{align}
    \zeta U_{n,s} - U_{n+1,s} = \frac{n\hat{\Phi}}{2} M_{n-1,s},
    \label{eq:moment}
\end{align}
where $\zeta = (\omega/k-u_s)/v_{ts}$, $\hat{\Phi} = q_s\Phi/T_s$, and $T_s = m_s v_{ts}^2/2$ is the effective temperature. 

The kinetic response function, $R_s(\zeta) = -U_{0,s}/\hat{\Phi}$, 
can be evaluated from \eqref{eq:Vlasov},
\begin{align}
    R_s(\zeta) = -\frac{v_{ts}}{2n_s} \int dw \, \frac{f_{0s}}{(w-\zeta)^2}.
    \label{eq:R_def}
\end{align}
In the case of a Maxwellian background, $R_s(\zeta) = -Z_s'(\zeta)/2 = 1 + \zeta Z_s(\zeta)$, where $Z_s$ is the plasma dispersion function. For damped modes, $R_s$ is defined by the analytic continuation of this integral from the upper half-plane (the Landau prescription). This requires $f_{0s}$ to admit an analytic continuation off the real $w$ axis. For a Maxwellian this is automatic.
The class of Landau fluid closures \cite{hammett1990fluid,hunana2018new,sun2026wave} assumes a linear closure relation for the highest evolved moment $N$, $U_{N,s} = \sum_{i<N} a_i U_{i,s}$, equivalent to replacing the kinetic response function by a two-point Pad\'{e} approximant, denoted $R_s^{n,n'}(\zeta)$. Here $n$ is the number of poles and $n'$ is the number of terms matched in the large-$\zeta$ expansion, with the remainder matched to the small-$\zeta$ expansion \cite{hunana2018new}.

While kinetic physics is included only approximately with a Pad\'{e} approach, no spurious instabilities are introduced in the process. For example, if we consider kinetic ions and adiabatic electrons, the quasineutrality condition reads
  $R^{n,n'}_s(\zeta) + T_i/T_e = 0$,
which agrees with the kinetic dispersion relation with the replacement $R_s(\zeta) \rightarrow R_s^{n,n'}(\zeta)$. Thus, any spurious roots arise only from the Pad\'{e} approximation itself. In particular, $R_s^{n,n'}$ can be chosen to be analytic in the upper half-plane, so that instabilities only arise from zeros of the dispersion relation, not from poles of the response function. Such response functions preserve causality \cite{nussenzveig1972causality}.

We propose an alternative closure strategy which, like Landau fluid closures, relies on the linearized moment equations in Fourier space, but instead allows for nonlinearity in the closure relation, $U_{N,s} = f(U_{i<N,s})$. In doing so, we retain the \textit{exact} linear physics of an eigenmode rather than relying on asymptotic expansion, retaining the complex analyticity of the response function.

The general strategy is to 1) estimate the mode frequency from the $n = 0$ moment equation \eqref{eq:moment}, 
\begin{align}
    \hat{\zeta} = \frac{U_{1,s}}{U_{0,s}},
    \label{eq:zeta_hat}
\end{align}
and 2) use the moment hierarchy to express the ratio $U_{n,s}/U_{0,s}$ as a function of $\zeta$. As an example, truncating at the $N = 2$ level yields the closure relation,
\begin{align}
    \frac{U_{2,s}}{U_{0,s}} = \zeta^2 + \frac{1}{2R_s(\zeta)}.
    \label{eq:beta_vlasov}
\end{align}
Here $\zeta$ and $\hat{\zeta}$ play distinct roles. The former is a parameter labeling a single eigenmode, while the latter is a functional of the instantaneous moments. The two agree exactly on an eigenmode. 

The closed fluid system is obtained by evaluating \eqref{eq:beta_vlasov} at the estimate $\hat{\zeta}$ and inverse Fourier transforming in time, with $\hat{\Phi}$ constrained by Poisson's equation:
\begin{align}
\left\{ \begin{array}{l}
    \frac{1}{v_{ts}}\left(\partder{}{t} + iku_s\right)U_{0,s} + ikU_{1,s} = 0 \\[4pt]
    \frac{1}{v_{ts}}\left(\partder{}{t} + iku_s\right)U_{1,s}
    + ik U_{0,s}\left(\hat{\zeta}^2 + \frac{1}{2R_s(\hat{\zeta})} \right) = - \frac{ik\hat{\Phi}}{2}
    \end{array}
    \right. .
\label{eq:closure_time_domain}
\end{align}
The spatial inverse Fourier transform must generally be performed numerically (similar to other Landau fluid closures \cite{dimits2014fast} such as Hammett-Perkins).
The estimate \eqref{eq:zeta_hat} is undefined where $U_{0,s}$ vanishes, which
occurs off of an eigenmode. Examples include interference nodes of superposed eigenmodes and initialization. 
The closure itself remains well behaved away from the negative imaginary $\hat{\zeta}$ axis, where the large-argument expansion of $Z_s$ applies. There, as $|\hat{\zeta}| \rightarrow \infty$, \eqref{eq:beta_vlasov} approaches
$3M_{2,s}$, the linearized adiabatic closure, so $U_{2,s}$
vanishes smoothly with $U_{0,s}$.
In practice, we clip $|\hat{\zeta}|$ at a value outside the mode region of
interest. For a single eigenmode, $\hat{\zeta}$ is constant in time and lies well within the clipped value, so the reported growth rates are unaffected. On the other hand, the nodes of a Landau-damped standing wave drive $\hat\zeta$ toward $-i\infty$. We examine this
case in the Supplemental Material \cite{supp}.

Since \eqref{eq:closure_time_domain} enforces the exact kinetic response on the eigenmode manifold, the fluid and kinetic dispersion relations coincide.
However, exactness on the eigenmode manifold does not by itself preclude spurious modes. Substituting a single-mode ansatz into the closed fluid equations shows that the fluid dispersion relation factorizes into the kinetic dispersion relation multiplied by a prefactor polynomial $P_N(\zeta)$. For the closures presented here, $P_2 = \zeta\zeta_* - \frac{1}{2}$ for the ITG closure, reducing to $P_2 = -\frac{1}{2}$ in the Vlasov limit. Only upper-half-plane zeros of $P_N$ would constitute spurious instabilities. Since $P_2$ has no upper-half-plane zeros, no spurious instabilities arise at $N=2$. 

We emphasize that the EKR closure is exact \textit{on the eigenmode manifold}. However, due to its nonlinearity, it does not satisfy the principle of superposition. Thus, if the moments represent a superposition of eigenmodes, the estimate $\hat{\zeta}$ does not correspond to any single eigenmode frequency, and the closure is no longer exact. 
We address this error with a neural-network closure below. 





\hd{Model problems}We begin with analysis of a bump-on-tail (BoT) instability, in which a Langmuir wave supported by the bulk is driven unstable by an energetic electron tail. The bulk electron population, with density $n_0$ and flow velocity $u$, is a cold fluid with linearized momentum equation, 
$\partial u/\partial t = - e E/m_e.$
The beam electron population, $f_b$, is described by \eqref{eq:Vlasov} and coupled with the bulk Maxwellian through Amp\`{e}re's law,
\begin{align}
    \epsilon_0 \partder{E}{t} = e n_0 u - j_b,
\end{align}
where $j_b = -e \int dv \, f_{1b} v = - e n_{b}\left(v_{tb}U_{1,b} +u_bU_{0,b}\right)$. The resulting wave equation describes Langmuir waves that are kinetically modified by the beam:
\begin{align}
    \partder{^2 E}{t^2} = - \omega_{p,e}^2 E + \frac{e n_b}{\epsilon_0} \left(v_{tb} \partder{U_{1,b}}{t} + u_b \partder{U_{0,b}}{t}\right). 
    \label{eq:modified_Langmuir}
\end{align}
Here $\omega_{p,e}^2 = n_0 e^2/(\epsilon_0 m_e)$ is the squared bulk plasma frequency. Equation \eqref{eq:modified_Langmuir} is evolved with the closed beam fluid equations \eqref{eq:closure_time_domain}. The beam is characterized by its relative density $\epsilon = n_b/n_0$ and normalized drift $u_b/v_{tb}$. 


We next consider the slab ITG instability. Following \cite{hammett1990fluid}, we model the ions using the linearized drift-kinetic equation,
\begin{align}
    \partder{f_{1i}}{t} + v_{\|} \hat{\bm{b}} \cdot \nabla f_{1i} + \bm{v}_E \cdot \nabla f_{0i}  + \frac{q_iE_{\|}}{m_i} \partder{f_{0i}}{v_{\|}} = 0,
\end{align}
where $\bm{v}_E$ is the $E \times B$ drift. The magnetic field is in the $\hat{\bm{z}}$ direction, with $f_{0i}$ a Maxwellian with moments depending on $x$. Defining the normalized variables $\zeta = \omega/(k_{\|} v_{ti})$, $w = v_{\|}/v_{ti}$, the Fourier-transformed equation reads:
\begin{align}
    - (\zeta - w)f_{1i} = \hat{\Phi}\left(\zeta_* \left[1 + \eta \left(w^2 - \frac{1}{2}\right)\right] - w\right)f_{0i},
\end{align}
where $\eta = L_{n}/L_T$ is the ratio of the ion density length scale to the temperature length scale, $\omega_* = (T_i/q_iB) k_y/L_n$ is the diamagnetic drift frequency, and $\zeta_* = \omega_*/(k_{\|} v_{ti})$. The system is closed by assuming quasineutrality with an adiabatic electron species, $U_{0,i} = \tau \hat{\Phi}$,
where $\tau = T_{0i}/T_{0e}$. The closure relation reads:
\begin{align}
    \frac{U_{2,i}}{U_{0,i}} = \zeta^2 - \frac{\zeta \zeta_* - \frac{1}{2}}{R_i(\zeta)},
\end{align}
where $R_i(\zeta) = -U_{0,i}/\hat{\Phi}$ is now the drift-kinetic ion response function, which depends on the drive parameters $\zeta_*$ and $\eta$ in addition to $\zeta$:
\begin{align}
    R_i(\zeta) = 1 - \eta \zeta \zeta_* + \left[\zeta - \zeta_*\left(1 - \frac{\eta}{2} + \eta \zeta^2\right)\right] Z(\zeta).
\end{align}
The closure recovers the Vlasov result \eqref{eq:beta_vlasov} in the limit $\zeta_* \rightarrow 0$. Furthermore, because the $n = 0$ moment equation acquires a source from the diamagnetic drive, $\zeta U_{0,i} - U_{1,i} = -\zeta_* \hat{\Phi}$, the frequency estimate \eqref{eq:zeta_hat} reads $\hat{\zeta} = U_{1,i}/U_{0,i} - \zeta_*/\tau$.


\hd{EKR validation}For both the BoT and ITG closures, the Vlasov-Poisson and drift kinetic ground truth is generated by discretizing velocity space with a uniform grid and using a Fourier spatial representation. The linear system is cast as an eigenvalue problem and time advanced by projecting the initial condition onto eigenmodes. The fluid system is also discretized with a Fourier spectral method and is time advanced using RK45. An initial amplitude for the electric field is applied, and fluid moments are initialized to zero. For the parameter sweep, the wavenumber at each grid point maximizes the kinetic growth rate, and growth rates are fit over a window spanning a fixed number of kinetic e-foldings.

\begin{figure}
\centering
\includegraphics[width=0.99\linewidth]{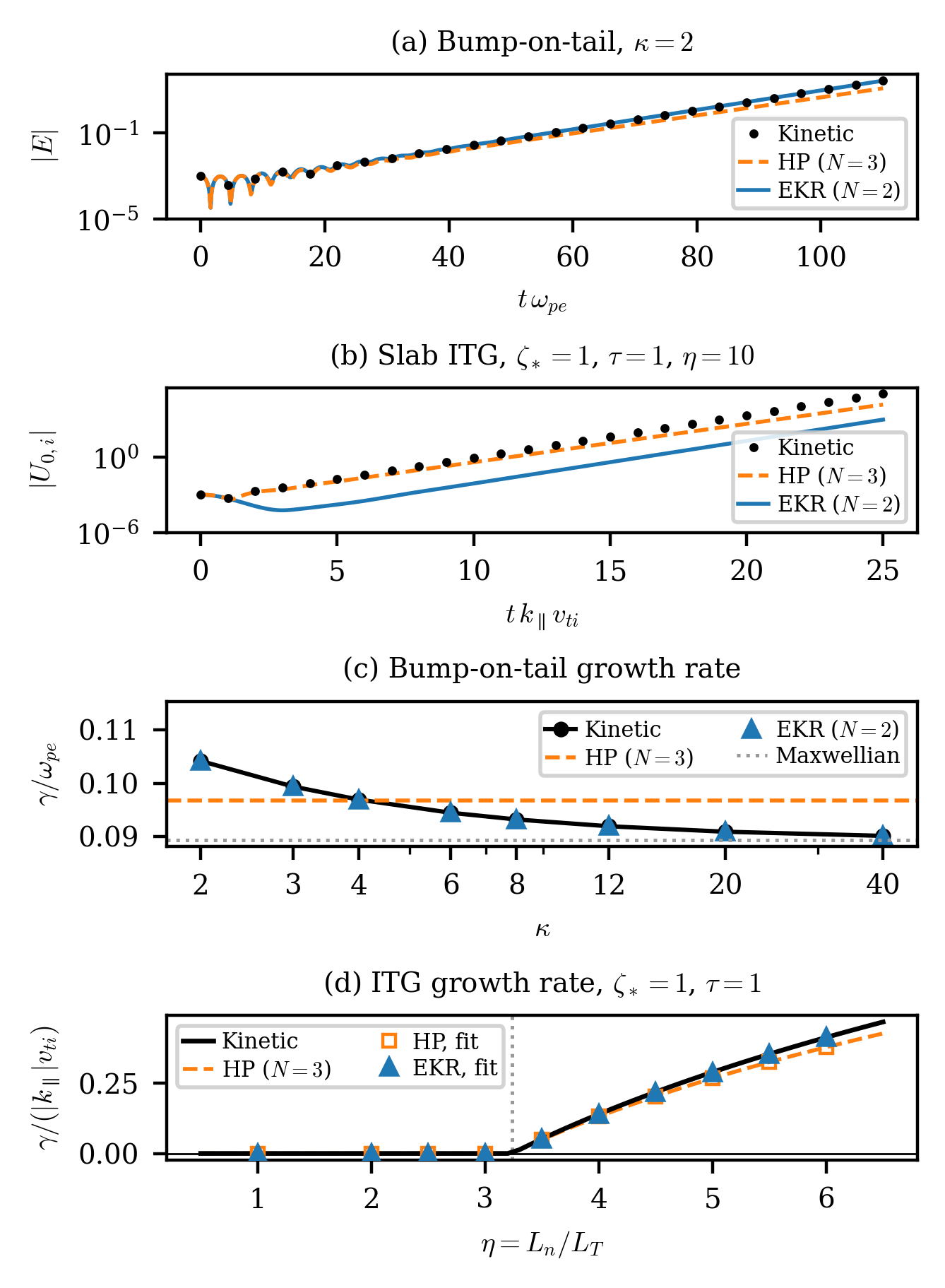}
\caption{Validation of the EKR closure. Bump-on-tail (a) field amplitude and (c) growth rate versus $\kappa$, for $u_b=4$, $\epsilon=0.02$, $k=0.295$. Slab-ITG (b) density amplitude and (d) growth rate versus $\eta$, with lines from the dispersion relations, markers from time-domain fits, and the dotted line at $\eta_{\rm th}=1+\sqrt{5}$.}
\label{fig:validation}
\end{figure}

In Fig.~\ref{fig:validation} we compare the evolution of the Hammett-Perkins (HP) and EKR closures with the kinetic ground truth. For generic initial conditions, the error in the growth rate computed from the HP closure can be as large as $\sim 10\%$, especially for intermediate values of $\zeta$ (e.g., $\zeta = 0.784+0.775i$ for the ITG example). The EKR closure reproduces the growth rate with an error of $2\times 10^{-5}$ and $3\times 10^{-5}$, for the BoT and ITG problems, respectively, arising due to competition from transient subdominant modes. This is visible in Fig.~\ref{fig:validation}(b), and we return to it below.


Figures \ref{fig:validation}(a) and \ref{fig:validation}(c) demonstrate the generality of the EKR closure with respect to the equilibrium distribution function. We replace the Maxwellian beam with a $\kappa$ distribution, $f_{0b} \propto (1 + w^2/b^2)^{-(\kappa+1)}$ with $b^2 = \kappa - 1/2$ chosen so that the density and temperature are unchanged, and only the shape of the tail differs. 
Since $f_{0b}$ is rational, the integral \eqref{eq:R_def} can be evaluated in closed form. The kinetic growth rate varies by $\sim 16\%$ over the range of $\kappa$ shown, and the EKR closure follows it throughout. The HP prediction is independent of $\kappa$ by construction. 

Figure \ref{fig:validation}(d) extends the ITG comparison across the temperature-gradient drive parameter at fixed $\zeta_* = 1$. The EKR closure reproduces the kinetic growth rate at every $\eta$, including the linear stability threshold $\eta_{\rm th} = 1 + \sqrt{5}$, while the HP closure also captures the threshold exactly but underestimates the growth rate above it.



\hd{Learned closures}Because of linearity of the Landau fluid closure relations, any initial condition cleanly decomposes into the fluid system's eigenmodes, and the superposition principle is automatically satisfied. On the other hand, the nonlinear EKR closure does not. Since the mode frequency is estimated from the instantaneous moment ratio $\hat{\zeta}$, an incorrect frequency will be deduced for a general superposition. The final growth rate prediction may still be accurate for unstable systems, since the fastest-growing mode will dominate at long enough times. However, as seen in Fig.~\ref{fig:validation}(b), the initial transient may be incorrect.

To close this gap, we train a NN on the mapping $\bm{U} = \{U_0, \dots, U_{N-1}, \hat{\Phi}\} \rightarrow U_N$. Linear physics is invariant under multiplication of $\bm{U}$ by a complex constant $a$, so the closure must be \textit{equivariant}, $U_N(a\bm{U}) = aU_N(\bm{U})$. We build this symmetry into the architecture directly \cite{villar2021scalars}. A multi-layer perceptron $\mathcal{N}_i$ acts only on the Hermitian matrix $G = \bm{U}\bm{U}^{\dagger}/|\bm{U}|^2$, whose entries $U_iU_j^*/|\bm{U}|^2$ are invariant under complex rescaling, and its outputs multiply the moments, $U_N = \sum_{i<N}\mathcal{N}_i(G)\,U_i$. 

An $M$-mode superposition is characterized by $2M-1$ complex parameters ($M$ frequencies and $M-1$ relative amplitudes), while the moment and potential vector supplies $N$ complex constraints ($N+1$ components modulo a common complex amplitude). Resolving $M$ modes therefore requires $N \ge 2M-1$. The EKR closure ($N=2$) resolves a single mode, and a two-mode superposition requires $N=3$. 

To generate the training data, two eigenmode parameters $\zeta_1$, $\zeta_2$ are drawn from a rectangle in the complex plane, $\mathrm{Re}\,\zeta \in (-3, 3)$, $\mathrm{Im}\,\zeta \in (-1, 1.5)$. The complex amplitudes are chosen on the unit sphere. For each eigenmode of frequency $\zeta$
the response function fixes $U_{0,s} = -R_s(\zeta)\hat\Phi$, and the
hierarchy \eqref{eq:moment} then generates every higher moment,
$U_{n+1,s} = \zeta U_{n,s} - (n/2)\hat\Phi M_{n-1,s}$. 

The same samples can instead be used to learn a \textit{linear} closure: we introduce the \textit{learned Pad\'{e}} closure, in which the coefficients of $U_{N,s} = \sum_{i<N} a_i U_{i,s}$ are fit by least squares over the same training data to the single-mode closure ratio, $\alpha_N(\zeta) \approx \sum_{i<N} a_i \alpha_i(\zeta)$, where $\alpha_j(\zeta) = U_{j,s}/U_{0,s}$ on an eigenmode of frequency $\zeta$. Whereas Hammett-Perkins and Hunana fix the coefficients by asymptotic matching at isolated points, the learned closure distributes its error over the region of the complex plane where the modes of interest reside. (A related strategy is employed in \cite{spong2013simulation}.) Because the closure remains linear in the moments, it retains the structural guarantees of the Landau fluid family.
Superposition holds exactly, the reality conditions can be imposed on the coefficients, and the fitted poles can be verified to lie in the lower half-plane, ensuring causality.
On the other hand, the NN reproduces the kinetic response more accurately, at the cost of exact superposition and causality. 
As shown in Fig.~\ref{fig:opt_hunana_sweep}, learning the coefficients reduces the median growth-rate error by a factor of $\sim 4$ relative to Hammett-Perkins at $N=3$ and $\sim 2$ relative to Hunana at $N=4$. The nonlinear EKR and $N=4$ NN closures remain one to two orders of magnitude more accurate. 

\begin{figure}
    \centering
    \includegraphics[width=1.0\linewidth]{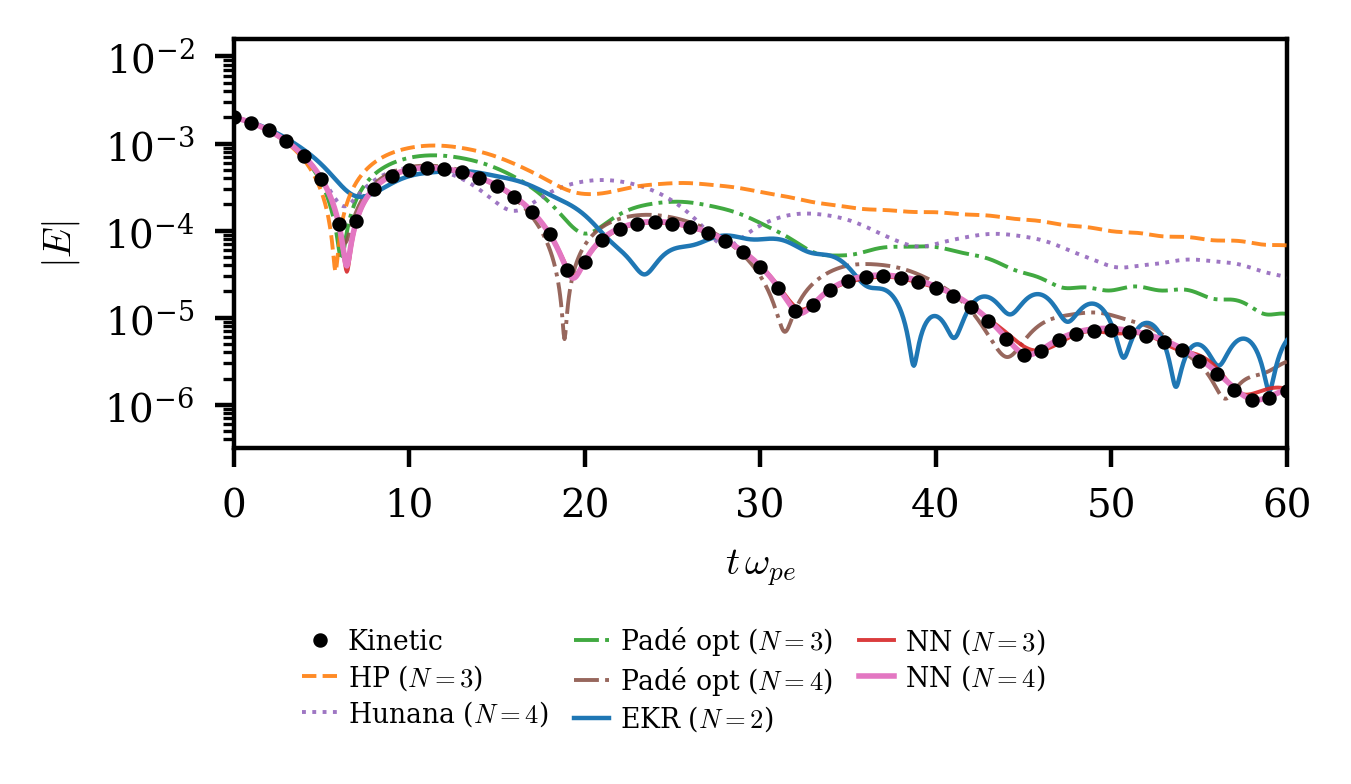}
    \caption{Superposition of two Landau-damped eigenmodes of the bump-on-tail problem, $u_b = 1.5$, $k = 0.40$, $\epsilon = 0.05$, $\omega_{1,2} = 1.205 - 0.119i$, $0.717 - 0.106i$.}
    \label{fig:damped_super}
\end{figure}

\begin{figure*}
\includegraphics[width=0.75\linewidth]{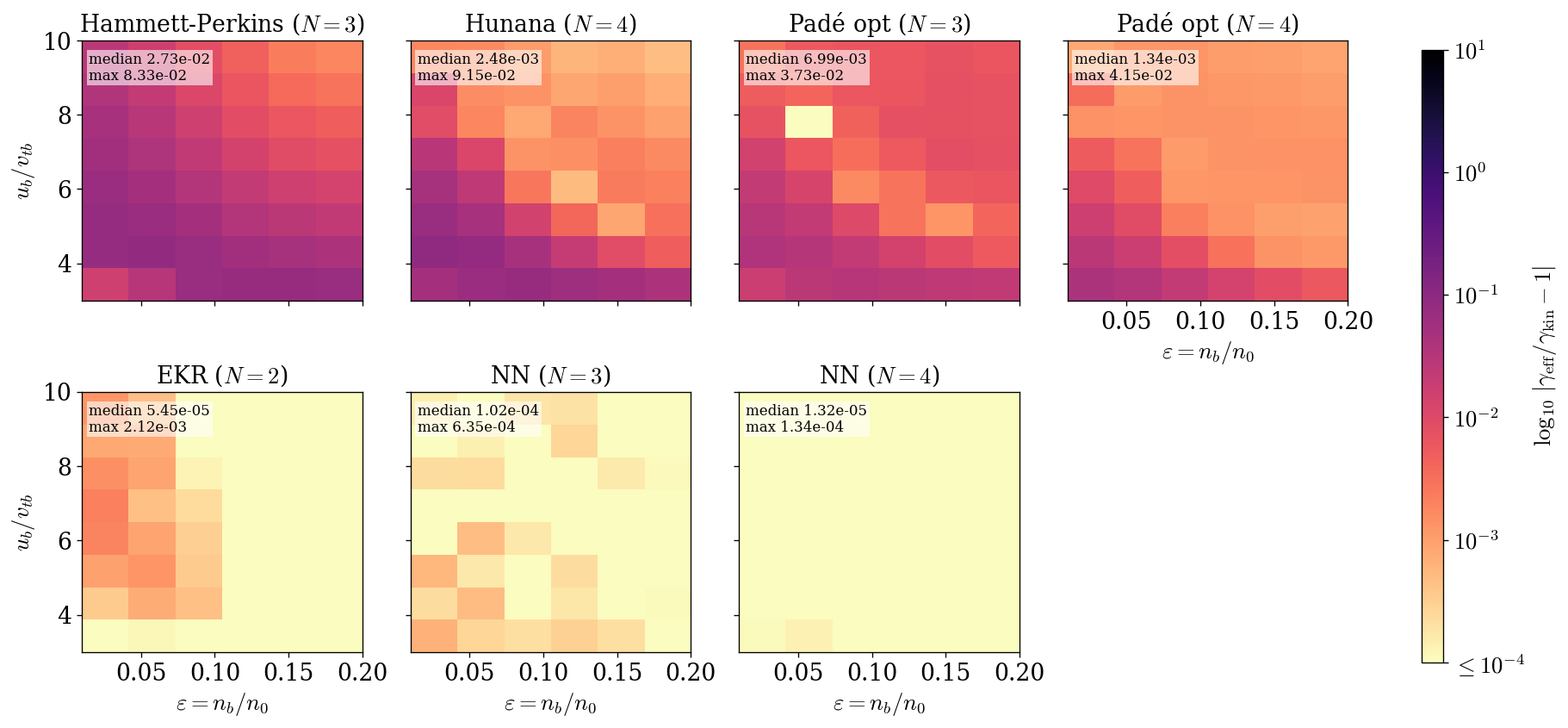}
\caption{Relative growth-rate error $|\gamma_{\rm eff}/\gamma_{\rm kin}-1|$ over the bump-on-tail parameter space for the seven closures of Fig.~\ref{fig:damped_super}. Insets give the median and maximum over the grid.}\label{fig:opt_hunana_sweep}
\end{figure*}

In Fig.~\ref{fig:damped_super}, we consider a superposition of two Landau-damped eigenmodes. The $N=3$ and $N=4$ NN closures reproduce the kinetic evolution most closely, consistent with the bound $N \ge 2M-1$, while the $N=2$ EKR closure latches onto a single ``effective'' mode and misrepresents the beat structure. Among the linear closures, which satisfy superposition exactly, the learned $N=4$ Pad\'{e} closure performs best, following the beat envelope, while the asymptotically matched Hammett-Perkins and Hunana closures substantially underestimate the damping. The Landau-damped standing wave is likewise reproduced by the $N=4$ NN closure \cite{supp}.

Figure \ref{fig:opt_hunana_sweep} extends the comparison to the unstable branch, over a grid in the beam drift $u_b$ and relative density $\epsilon$. Since the two-point Pad\'{e} approximants are matched in the small- and large-$|\zeta|$ limits, their largest errors occur at intermediate $|\zeta|$. Increasing either $u_b$ or $\epsilon$ increases $|\zeta|$, so the linear closures are least accurate along the small-$u_b$ edge, and the learned coefficients improve accuracy across the grid. The EKR and $N=4$ NN closures reach median errors of $5\times 10^{-5}$ and $1\times 10^{-5}$, comparable to the transient contamination of the growth-rate fit itself. The $N=3$ NN closure, although an order of magnitude less accurate than the $N=4$, still outperforms every linear closure.

\hd{Discussion}For both model problems, and for $\kappa$-distributed as well as Maxwellian equilibria, the nonlinear EKR closure reduces the growth-rate error from the $\sim 10\%$ level of Hammett-Perkins to $\sim 10^{-5}$, and reproduces the ITG stability threshold exactly. The EKR, learned Pad\'{e}, and NN closures extend to any complex analytic non-Maxwellian equilibrium.

We conclude that a closure evaluated on the instantaneous moments
$U_j$, $j < N$ and the potential, can only represent an $M$-mode superposition for $2M \leq N + 1$.
This counting argument constrains any closure whose inputs are the instantaneous moments and field, including data-driven closures trained on kinetic simulations (e.g., \cite{cheng2023data,Huang2025MLHeatFluxClosure,wei2023data}). A closure with access to time history can, in principle, resolve correspondingly more modes. 

Equivariance builds the symmetry of linear physics into the NN closure exactly, but not causality. The poles of the linear learned Pad\'{e} closure can be verified directly, whereas the NN carries no such guarantee.
Imposing the causality conditions during training \cite{vemuri2025rampinn}
or training the closure in-the-loop by differentiating through the fluid solver \cite{um2020solver,kochkov2024neural} offer routes to avoiding spurious instabilities.


All examples discussed are at a single wavenumber, but are trivially generalized to broadband initial conditions.
The moments, and hence the frequency estimate $\hat{\zeta}$, are functions of $k$, and the closure is evaluated independently at each wavenumber. The nonlinearity of the EKR and NN closures acts on the moments at fixed $k$, so wavenumbers remain decoupled, and a NN trained on the moment hierarchy requires no retraining.


Finally, while the closures developed here are built from linearized physics,
the data-generation strategy is not restricted to it. The linearized hierarchy supplies exact samples densely covering the complex frequency
plane, and can regularize nonlinear training data in the regions where simulation sampling may be sparse.

\hd{Acknowledgments}This work was supported under grants from the National Science Foundation under award 2512171, by IFE COLoR under U.S. Department of Energy Grant No. DE-SC0024863, and by the U.S. Department of Energy under contracts DE-AC02-09CH11466 (StellFoundry), DE-SC0024548 (HiFiStell), and DE-SC0024630. Part of this research was performed while the authors were visiting the Institute for Pure and Applied Mathematics (IPAM), which is supported by the National Science Foundation (Grant No. DMS-1925919/2422832). We thank Anna Tenerani and Ben Zhu for illuminating discussions. The authors used Claude Code (Anthropic) between May and September 2026 to assist in software development, including developing the driver and analysis scripts used for the parameter scans. Claude code was also used to suggest revisions to the prose. All code and text were reviewed and edited by the authors, who take full responsibility for the content, including all references.

\hd{Data availability}The code and data that support the findings of
this article are openly available at Ref.~\cite{bumpontail2026}.

\bibliography{apssamp}

\ifdefined\arxivversion
\clearpage
\onecolumngrid
\begin{center}
\textbf{Supplemental Material for ``Nonlinear kinetic closures for linear instabilities''}
\end{center}
\setcounter{figure}{0}
\setcounter{table}{0}
\renewcommand{\thefigure}{S\arabic{figure}}
\renewcommand{\thetable}{S\arabic{table}}

\section*{Landau-damped standing wave}

As a test of the closures away from the eigenmode manifold, we consider a single Maxwellian electron species with a neutralizing ion background at $k\lambda_D = 0.4$, where $\lambda_D = v_{te}/(\sqrt{2}\,\omega_{pe})$ is the Debye length, initialized with a pure density perturbation, $f_1(v,0) = \epsilon f_0(v)$, so that $U_n(0) = \epsilon M_n$. Poisson's equation constrains the potential as $\hat\Phi = U_0/(k\lambda_D)^2$. The closed moment systems are otherwise those of the main text. The kinetic solution is a Langmuir oscillation Landau damped at the least-damped root, $\omega/\omega_{pe} = \pm 1.285 - 0.066i$. $U_0(t)$ is a real standing wave that passes through zero twice per period. This is a two-mode superposition ($M=2$), for which the counting argument requires $N \geq 3$.

At each node $U_0 = 0$ while $U_1 \neq 0$, so the EKR frequency estimate $\hat\zeta = U_1/U_0$ diverges. Moreover, since $U_0$ is real and $U_1 = (i/k)\,\partial_t U_0$ is imaginary, $\hat\zeta$ is confined to the imaginary axis, where the dispersion relation has no root. The EKR closure therefore cannot lock onto the Landau root and collapses onto non-oscillatory decay (Fig.~\ref{fig:standing}). The linear closures oscillate but misplace the root, with frequency errors of 2--7\% and damping-rate errors of 5--60\% (Table~\ref{tab:standing}). The NN closures, whose inputs $G(\bm{U})$ remain well defined at the nodes, recover the frequency and damping rate to $\sim 1\%$ at $N=3$ and $0.1\%$ at $N=4$.

\begin{figure}[h!]
\centering
\includegraphics[width=0.98\linewidth]{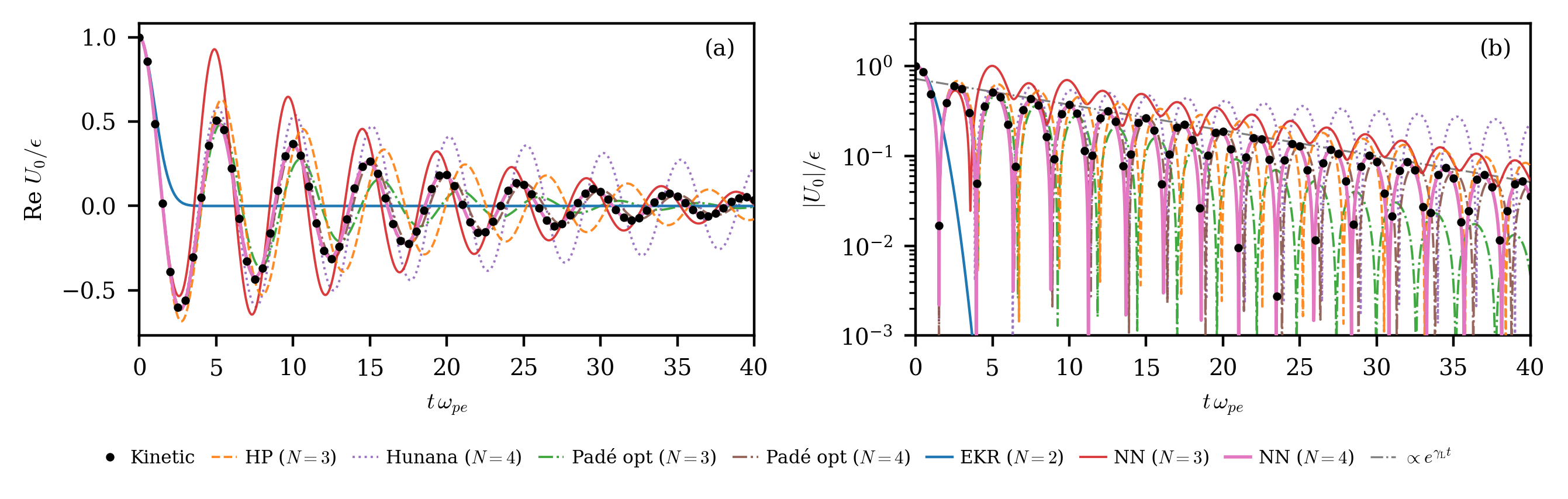}
\caption{Landau-damped standing wave at $k\lambda_D=0.4$: (a) density moment $\mathrm{Re}\,U_0/\epsilon$ and (b) its magnitude, for the kinetic solution and the seven closures of Fig.~2 of the main text. The gray line marks the Landau damping rate $\gamma_{\rm L}$.}
\label{fig:standing}
\end{figure}

\begin{table}[h!]
\caption{Relative errors in the frequency and damping rate for the standing wave of Fig.~\ref{fig:standing}, fit from the zero crossings and envelope of $\mathrm{Re}\,U_0$ for $t\,\omega_{pe} > 5$.}
\label{tab:standing}
\begin{ruledtabular}
\begin{tabular}{lcc}
Closure & $\delta\omega_r/\omega_r$ & $\delta\gamma/\gamma$ \\
\hline
Hammett--Perkins ($N=3$) & $-7.4\%$ & $-12.0\%$ \\
Hunana ($N=4$) & $-2.7\%$ & $-58.6\%$ \\
Learned Pad\'e ($N=3$) & $-5.6\%$ & $+61.4\%$ \\
Learned Pad\'e ($N=4$) & $-1.8\%$ & $+4.6\%$ \\
EKR ($N=2$) & \multicolumn{2}{c}{no oscillation} \\
NN ($N=3$) & $+1.1\%$ & $+0.5\%$ \\
NN ($N=4$) & $+0.0\%$ & $-0.1\%$ \\
\end{tabular}
\end{ruledtabular}
\end{table}

\fi

\end{document}